\documentclass[aps,prl,twocolumn,superscriptaddress,showpacs]{revtex4}
\usepackage{amssymb}
\usepackage{amsmath}
\usepackage{graphicx}

\begin{document}

\title{Lattice-Induced Double-Valley Degeneracy Lifting in Graphene by a
Magnetic Field }
\author{Igor A. Luk'yanchuk}
\affiliation{University of Picardie Jules Verne, Laboratory of Condensed Matter Physics,
Amiens, 80039, France}
\affiliation{L. D. Landau Institute for Theoretical Physics, Moscow, Russia}
\author{Alexander M. Bratkovsky}
\affiliation{Hewlett-Packard Laboratories, 1501 Page Mill Road, Palo Alto, California
94304}
\date{\today }

\begin{abstract}
We show that the recently discovered double-valley splitting of the
low-lying Landau level(s) in the Quantum Hall Effect in graphene can
be explained as \textit{perturbative} orbital interaction of intra-
and inter-valley microscopic orbital currents with a magnetic field.
This effect is provided by the translational-non-invariant terms
corresponding to graphene's crystallographic honeycomb symmetry but
do not exist in the relativistic theory of massless Dirac Fermions
in Quantum Electrodynamics. We discuss recent data in view of these
results.
\end{abstract}

\pacs{71.70.-d,73.43.-f, 81.05.Uw}
\maketitle

The recent discovery of massless charge carriers with linear conic spectrum
- Dirac fermions (DF) in both graphite \cite{Lukyanchuk2004} and graphene
\cite{Novoselov2005,Zhang2005} has prompted researchers to revisit many
basic ideas in Solid State Physics based on relativistic particle physics on
a lattice. The existence of Dirac fermions has been confirmed by direct
ARPES \cite{Zhou,Eli} and STS \cite{Andrei} measurements and by Quantum Hall
Effect (QHE) measurements in both graphite \cite%
{Kopelevich2003,Lukyanchuk2006} and graphene \cite{Novoselov2005,Zhang2005},
It has been recognized that the analogy between DF\ in graphene and
relativistic massless DF in Quantum Electrodynamics (QED) can be used
fruitfully to explore the properties of graphene-based systems \cite%
{Katsnelson2007}. One of the most important consequences of this analogy is
the peculiar quantization of relativistic Landau Levels (LLs) in a magnetic
field that are symmetric for positive (electron) and negative (hole)
energies. The LL\ energies are proportional $\pm \sqrt{n}$ as a function of
the level number $n$ away from the zero LL ($n=0)$ that is positioned
exactly at the Dirac point where electron and hole spectra touch at zero
magnetic field, $\boldsymbol{H}=0$. In addition to the infinite Landau
degeneracy, each level is double spin-degenerate, which is a direct
consequence of the relativistic (Lorenz) invariance of the QED equations.

The QED double-spin degeneracy corresponds to the double-valley degeneracy
of LLs in graphene and, together with conventional spin-degeneracy (not
considered by the QED analogy), produces the conductance steps of $4e^{2}/h$
\cite{Gusynin2005} observed in the semi-integer QHE \cite%
{Novoselov2005,Zhang2005}, double the size of standard steps. Importantly,
therefore, the recently discovered double-valley splitting for (at least)
the zero Landau level \cite{Zhang2006,Abanin2007} indicates a breakdown of
relativistic invariance in graphene. Several mechanisms \cite{Yang2007}
based on \emph{spontaneous} symmetry breaking driven by either long-range
Coulomb interaction \cite%
{Khveshchenko2001,Alicea2006,Gusynin1994,Fertig2006,Gusynin2006},
field-enhanced electron-phonon interaction \cite{Fuchs2007}, \ disorder \cite%
{Nomura2006,Goerbig2006,Abanin2007bis} or edge effects \cite%
{Abanin2007,Castro2006, Abanin2006} have been proposed to explain this
phenomenon.

In this paper, we demonstrate that the valley gap opening for low-lying LLs
is the \emph{intrinsic}\textit{\ }property of graphene-like systems. These
systems have a honeycomb crystallographic group that is different from the
relativistic Lorenz group in QED albeit resulting in a similar Dirac-like
equation for non-interacting fermions in zero magnetic field, $\boldsymbol{H}%
=0$. The difference becomes apparent in an applied magnetic field
when the additional translational-non-invariant terms accounting for
interaction of microscopic intra- and inter-valley orbital currents
with the magnetic field appear in the graphene Hamiltonian. The
effect of the double-valley LL
splitting has, therefore, a much more natural explanation as a \emph{%
perturbative} non-critical orbital splitting that is of the same order as
the standard Zeeman spin-splitting.

We first consider the origin and symmetry properties of the Hamiltonian, the
spectrum, and the wave functions of conducting electrons (holes) in the
vicinity of two crystallographically nonequivalent opposite corners $%
\boldsymbol{K}_{1,2}$ (also denoted as $\boldsymbol{K}$ and $\boldsymbol{K}%
^{\prime }$) of the hexagonal Brillouin Zone of graphene \emph{at zero field}%
, $\boldsymbol{H}=0$. The wave functions of the zero-energy states are
located exactly at $\boldsymbol{K}_{1,2}$ and can be linearly expanded over
a 4-component Bloch basis (irreducible representation) of the K-point \cite%
{Landau5}:
\begin{equation}
\widetilde{\Psi }\equiv \{\Psi _{i}\}_{i=1-4}=\{\Psi _{K_{1}}^{A},\Psi
_{K_{1}}^{B},\Psi _{K_{2}}^{B},\Psi _{K_{2}}^{A}\}^{T}.
\end{equation}%
(for symmetry reasons our set $\{\Psi _{i}\}$ is different from the commonly
used $\{\Psi _{K_{1}}^{A},\Psi _{K_{1}}^{B},\Psi _{K_{2}}^{A},\Psi
_{K_{2}}^{B}\}$).

\begin{figure}[t]
\centering
\includegraphics [width=4.4cm] {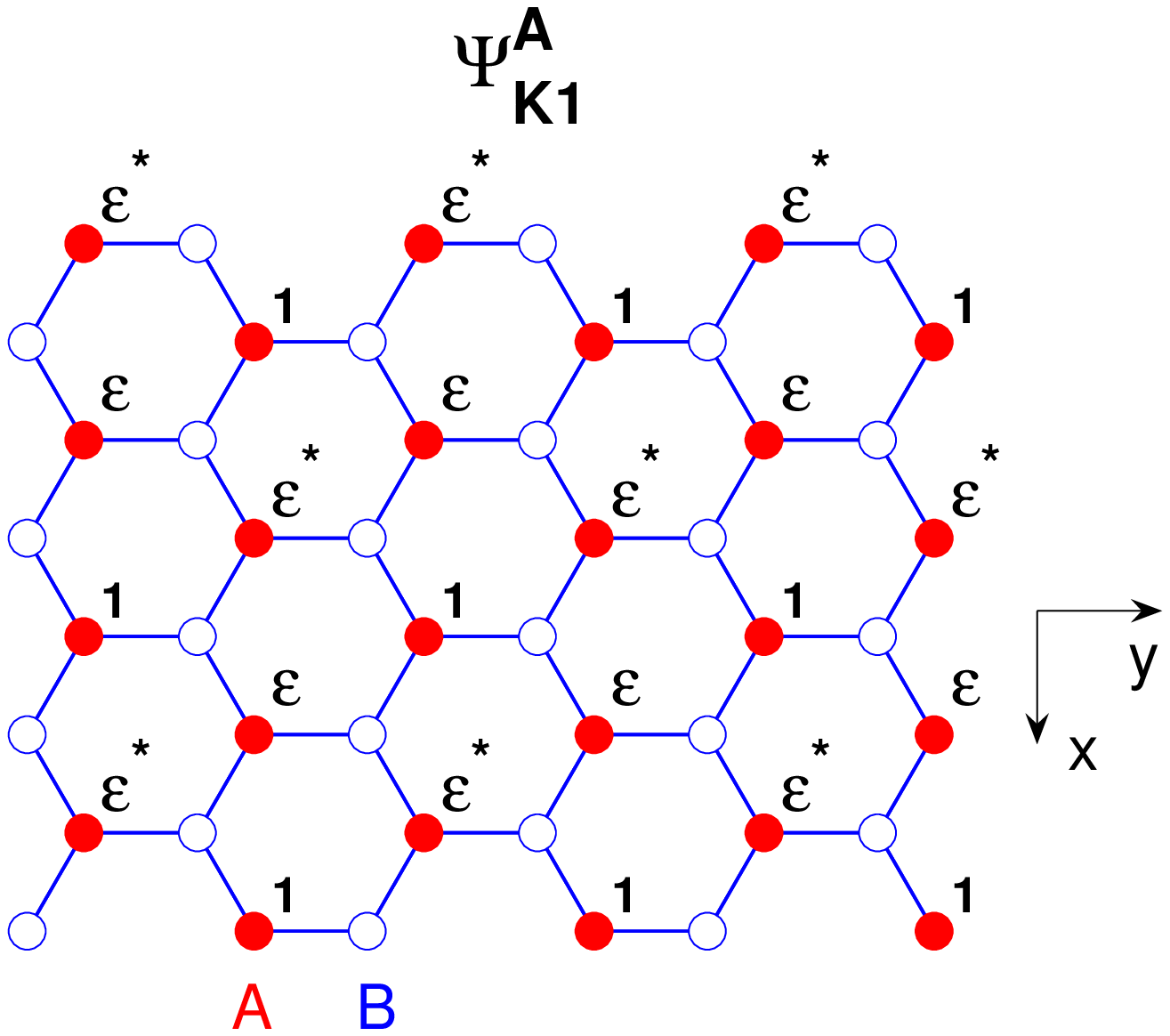} \ \
\includegraphics
[width=3.7cm]{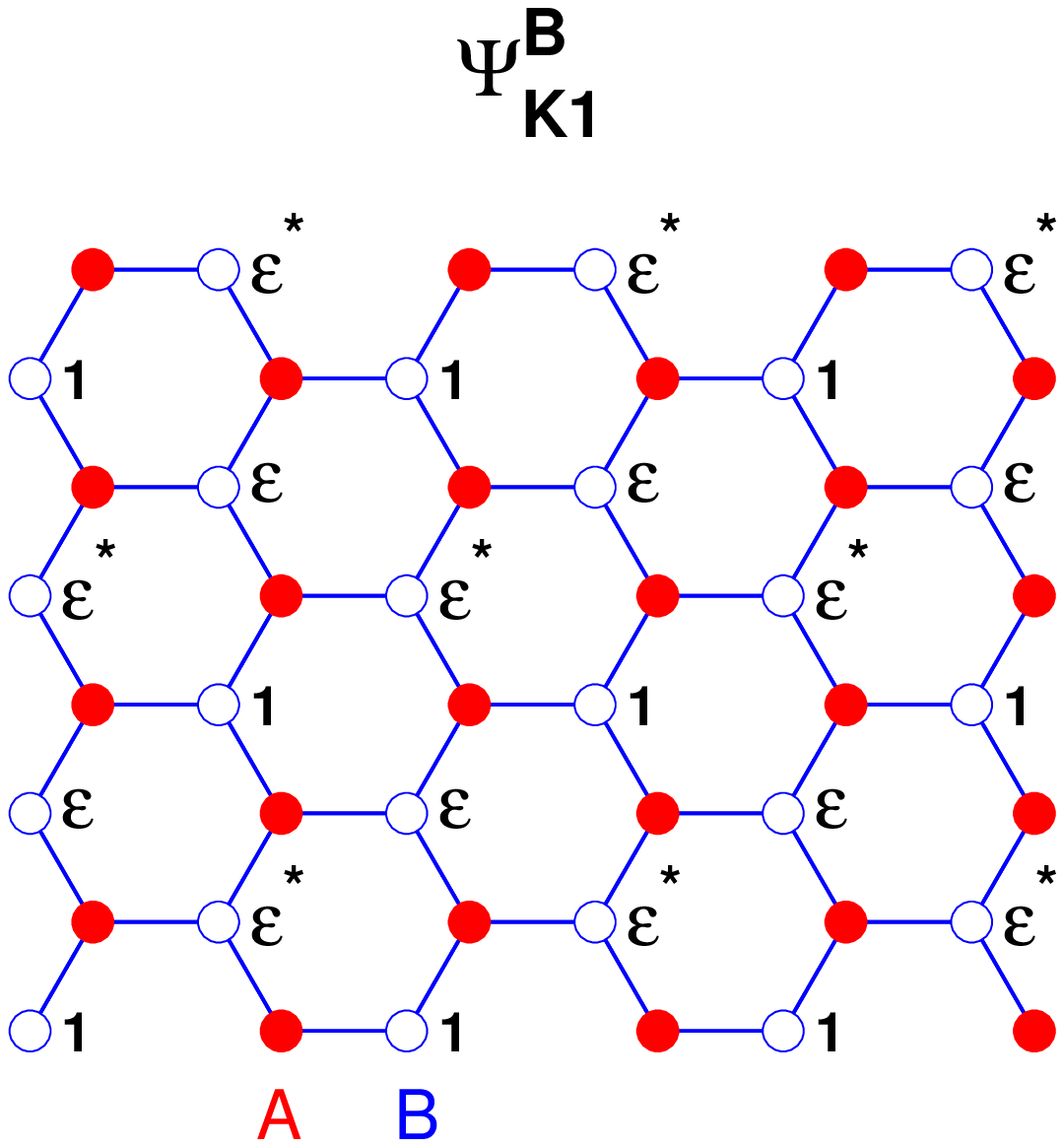} \newline \caption{$K_{1}$-point Bloch
functions $\Psi _{K_{1}}^{A}$ and $\Psi _{K_{1}}^{B}$. The
$K_{2}$-point Bloch functions $\Psi _{K_{2}}^{A}$ and $\Psi
_{K_{2}}^{B}$ are obtained from them by applying complex
conjugation.} \label{Bloch}
\end{figure}

\begin{table}[b]
\caption{Transformation properties of the Bloch spinor $\widetilde{\Psi }$ ($%
\protect\varepsilon =e^{2\protect\pi i/3}$)}
\label{TabSym}\centering \vspace{5mm}
\begin{tabular}{lcccc}
\hline\hline \null \ \ \ \ \ \ \ \ \  & \ \ \ $C_{6}$ \ \ \  & \ \ \
$\sigma _{y}$ \ \ \ & \ \ \ $\sigma _{x}$ \ \ \  & \ \ \
$\widehat{T}_{{12}}$ \ \ \  \\ \hline $\Psi _{K_{1}}^{A}$ &
$\overline{\varepsilon }\Psi _{K_{2}}^{B}$ & $\Psi
_{K_{2}}^{A}$ & $\Psi _{K_{1}}^{B}$ & $\varepsilon \,\Psi _{K_{1}}^{A}$ \\
$\Psi _{K_{1}}^{B}$ & $\varepsilon \Psi _{K_{2}}^{A}$ & $\Psi _{K_{2}}^{B}$
& $\Psi _{K_{1}}^{A}$ & $\varepsilon \,\Psi _{K_{1}}^{B}$ \\
$\Psi _{K_{2}}^{A}$ & $\varepsilon \,\Psi _{K_{1}}^{B}$ & $\Psi _{K_{1}}^{A}$
& $\Psi _{K_{2}}^{B}$ & $\overline{\varepsilon }\,\Psi _{K_{2}}^{A}$ \\
$\Psi _{K_{2}}^{B}$ & $\overline{\varepsilon }\Psi _{K_{1}}^{A}$ & $\Psi
_{K_{1}}^{B}$ & $\Psi _{K_{2}}^{A}$ & $\overline{\varepsilon }\,\Psi
_{K_{2}}^{B}$ \\ \hline\hline
\end{tabular}%
\end{table}
It is the transformation properties of spinor-like function $\widetilde{\Psi
}$ under the action of the graphene crystallographic group
\begin{equation}
G=\{C_{6},C_{3},C_{2},\sigma _{x},\sigma _{y},R\}\times \{\boldsymbol{T}_{1},%
\boldsymbol{T}_{2}\},
\end{equation}%
(Table \ref{TabSym}) that define all the physical properties of charge
carriers in graphene. Here, $\boldsymbol{T}_{1,2}$ are the lattice
translations; other notation are the same as in \cite{Landau5}. The physical
properties can be obtained either directly from the standard Tables of
Irreducible Representations of Crystallographic Groups \cite{Kovalev} or
from the explicit form of $\widetilde{\Psi }$ in a tight-binding
approximation of the carbon $p_{z}$ orbitals marked as $\pi (\mathbf{r})$
(see also Fig.~\ref{Bloch}):
\begin{equation}
\Psi _{K_{1,2}}^{A(B)}=e^{\frac{2}{3}s_{A(B)}i\pi }\sum_{nm}e^{s_{K_{1,2}}%
\frac{2}{3}i\pi (n+m)}\pi (\boldsymbol{r}-\boldsymbol{t}_{nm}^{A(B)})
\end{equation}%
where $s_{A(B)}=+(-)1$, $s_{K_{1,2}}=\pm 1$ and $\mathbf{t}_{nm}^{A(B)}$ are
the $A$ ($B)$ sublattice coordinates.

Wave functions of states deviating from $\boldsymbol{K}_{1,2}$ by a small
vector $\boldsymbol{k}=(k_{x},k_{y})$ can also be expanded over the basis $%
\Psi _{i}(r)$, but with slowly space-varying envelopes $\widetilde{F}^{%
\mathbf{k}}\equiv F_{i}^{\mathbf{k}}(r)$:
\begin{equation}
\Phi ^{\mathbf{k}}(r)=\sum_{i=1}^{4}F_{i}^{\mathbf{k}}(r)\Psi _{i}(r).
\label{Exp}
\end{equation}%
The energy spectrum $E(\mathbf{k})$ and the corresponding envelope functions
$\widetilde{F}^{\mathbf{k}}(r)$ are the eigenvalues and eigenfunctions of
the usual $\boldsymbol{Kk}$-perturbation equation:
\begin{equation}
\widehat{H}\ \widetilde{F}^{\mathbf{k}}(r)\ =\ E(\boldsymbol{k})\ \widetilde{%
F}^{\mathbf{k}}(r)
\end{equation}%
where the $\boldsymbol{Kk}$-perturbation Hamiltonian,
\begin{equation}
\widehat{H}=v\left(
\begin{array}{cccc}
0 & \widehat{k}_{x}+i\widehat{k}_{y} & 0 & 0 \\
\widehat{k}_{x}-i\widehat{k}_{y} & 0 & 0 & 0 \\
0 & 0 & 0 & -\widehat{k}_{x}+i\widehat{k}_{y} \\
0 & 0 & -\widehat{k}_{x}-i\widehat{k}_{y} & 0%
\end{array}%
\right)   \label{Dir}
\end{equation}%
(with $\widehat{\boldsymbol{k}}=-i\hbar \boldsymbol{\nabla }$) was obtained
as a most general $4\times 4$ matrix that is linear in $k$ and conserves the
form $\langle \widetilde{\Psi }\widehat{H}\widetilde{\Psi }\rangle $ under
the action of the group $G$.

The Hamiltonian (\ref{Dir}) has the structure of the relativistic Dirac
Hamiltonian for massless fermions with a linear conical spectrum (Fig. \ref%
{FigLevels}a):
\begin{equation}
E(\boldsymbol{k})=\pm v|\boldsymbol{k}|,  \label{spectr}
\end{equation}%
and the corresponding system of eigenfunctions $\widetilde{F}^{\mathbf{k}}(r)
$ that is a linear superposition (with arbitrary complex constants $c_{1}$, $%
c_{2}$) of two-valley plane-wave functions:
\begin{equation}
\widetilde{F}^{\mathbf{k}}(r)=c_{1}\{\pm 1,e^{i\theta },0,0\}e^{i\mathbf{kr}%
}+c_{2}\{0,0,\pm 1,e^{i\theta }\}e^{i\mathbf{kr}},
\end{equation}%
where $\theta =\arctan \left( k_{x}/k_{y}\right) $ and the $\pm $ sign
corresponds to the upper (lower) branch of the conical spectrum (\ref{spectr}
).

Note, however, that the similarity exploited above with DF in QED is valid
only in the vicinity of points $\boldsymbol{K}_{1,2}$ and is only
approximate. In reality, the transformation properties of $\widetilde{\Psi }%
_{K}$ with respect to the graphene crystallographic group $G$ (Table \ref%
{TabSym}) are quite different from those of the real DF transforming with
respect to the Lorentz group. This provides the additional contributions to
QED-like terms, such as the triangular-wrapped nonlinear kinetic \cite%
{McCann2006} and relativistic-noninvariant Coulomb interaction terms \cite%
{Khveshchenko2001}.

\begin{figure}[t]
\centering
\includegraphics [width=7cm] {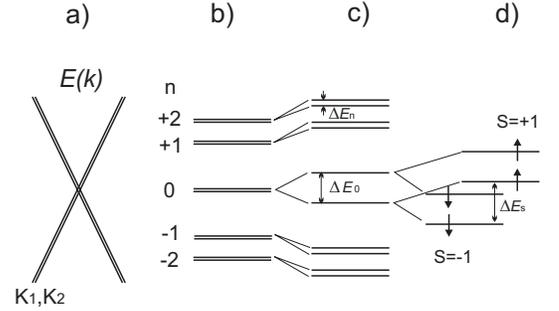}
\caption{(a) Two-valley degenerate Dirac-like spectrum $E(k)$ of charge
carriers in zero magnetic field, $H=0$. (b) Landau Level (LL) quantization
of Dirac Fermions. (c) Orbital valley splitting of LLs. (d) Additional
Zeeman spin-splitting of LLs (only the $n=0$ level is shown.)}
\label{FigLevels}
\end{figure}

The Bloch expansion (\ref{Exp}) also remains valid in a magnetic field $%
\boldsymbol{H,}$ although the slowly varying coefficients $\widetilde{F}(r)$
are now classified according to the discrete set of LLs (instead of
continuous $\boldsymbol{k}$). The usual way of introducing $\boldsymbol{H}=%
\boldsymbol{\nabla }\mathsf{\times }\boldsymbol{A}$ consists of the Peierls
substitution
\begin{equation}
\widehat{\boldsymbol{k}}\boldsymbol{\rightarrow }\widehat{\boldsymbol{k}}+%
\frac{\left\vert e\right\vert }{c}\boldsymbol{A,}  \label{Peierls}
\end{equation}%
which in the case of the Hamiltonian (\ref{Dir}) is the same as replacing $%
\widehat{k}_{x}\pm i\widehat{k}_{y}$ by the LL creation (annihilation)
operators $a^{\pm }$.

The Hamiltonian has a set of discrete LLs having a square-root energy
dependence on the level number $n=0,\pm 1,\pm 2,...$ in a magnetic field
(Fig \ref{FigLevels}b):
\begin{equation}
E_{n}=s_{eh}\sqrt{2v^{2}|e|\hbar H_{z}|n|/c},\qquad s_{eh}=\mathrm{sign}(n),
\label{Elevels}
\end{equation}%
which is quite different from the case of massive particles with $%
E_{n}=\hbar \omega _{c}(n+{\frac{1}{2}})\geq 0$, ($n=0,1,2...$), and has
solutions with values both above and below the zero-energy LL, $E_{0}=0$.
The corresponding eigenfunctions can be written as an expansion over the
n-th LL eigenfunctions $f_{n}(r)$:
\begin{eqnarray}
\widetilde{F}^{{n}}(r) &=&\{c_{1}f_{|n|}(r),s_{eh}ic_{1}f_{|n|-1}(r),  \notag
\\
&&c_{2}f_{|n|}(r),s_{eh}ic_{2}f_{|n|-1}(r)\},  \label{Flevels}
\end{eqnarray}%
[note that $f_{-1}(r)\equiv 0$]. Each level, including $n=0$, has the
two-valley degeneracy provided by the complex constants $c_{1}$,$c_{2}$, the
two-fold spin degeneracy and the infinite Landau degeneracy.

Although the Peierls substitution (\ref{Peierls}) conserves a
relativistic invariance of the Dirac equation in a magnetic field,
the discrete crystal lattice background leads to another,
\emph{weaker,} requirement that the Hamiltonian of the system should
be invariant with respect to the crystallographic group of graphene
in a magnetic field \cite{Landau8}:
\begin{equation}
G_{H}=\{C_{6}R,C_{3},C_{2}R,\sigma _{x}R,\sigma _{y}R\}.
\end{equation}%
In particular, this time non-invariant group $G_{H}\subset G$ does not
contain the translations $\boldsymbol{T}_{1,2}$ that are incompatible with
the translational magnetic group. The principal idea of the present work is
that the graphene Hamiltonian for charge carriers in magnetic field should
have the more general form:
\begin{equation}
\widehat{H}=\left(
\begin{array}{cccc}
\lambda \mu _{B}H_{z} & va^{+} & \gamma \mu _{B}H_{z} & 0 \\
va^{-} & -\lambda \mu _{B}H_{z} & 0 & -\gamma \mu _{B}H_{z} \\
\gamma \mu _{B}H_{z} & 0 & \lambda \mu _{B}H_{z} & -va^{+} \\
0 & -\gamma \mu _{B}H_{z} & -va^{-} & -\lambda \mu _{B}H_{z}%
\end{array}%
\right) ,  \label{HB}
\end{equation}%
($\mu _{B}=|e|\hbar /2mc$ ) that, besides the Peierls terms $a^{\pm }$,
contains the \textquotedblleft non-relativistic" $\lambda $- and $\gamma $-
corrections provided by the orbital interaction of Bloch electrons with the
magnetic field.

These terms keep $\langle \widetilde{\Psi }\widehat{H}\widetilde{\Psi }%
\rangle $ invariant under the operation of group $G_{H}$ and are produced by
the matrix elements:
\begin{gather}
\left\langle \Psi _{K_{1}}^{A}\widehat{V}\overline{\Psi }_{K_{2}}^{B}\right%
\rangle =-\left\langle \Psi _{K_{1}}^{B}\widehat{V}\overline{\Psi }%
_{K_{2}}^{A}\right\rangle   \label{gmatr} \\
=\left\langle \Psi _{K_{2}}^{B}\widehat{V}\overline{\Psi }%
_{K_{1}}^{A}\right\rangle =-\left\langle \Psi _{K_{2}}^{A}\widehat{V}%
\overline{\Psi }_{K_{1}}^{B}\right\rangle =\gamma \mu _{B}H_{z},  \notag
\end{gather}%
and
\begin{gather}
\left\langle \Psi _{K_{1}}^{A}\widehat{V}\overline{\Psi }_{K_{1}}^{A}\right%
\rangle =-\left\langle \Psi _{K_{1}}^{B}\widehat{V}\overline{\Psi }%
_{K_{1}}^{B}\right\rangle   \label{lmatr} \\
=\left\langle \Psi _{K_{2}}^{B}\widehat{V}\overline{\Psi }%
_{K_{2}}^{B}\right\rangle =-\left\langle \Psi _{K_{2}}^{A}\widehat{V}%
\overline{\Psi }_{K_{2}}^{A}\right\rangle =\lambda \mu _{B}H_{z},  \notag
\end{gather}%
of the perturbation operator
\begin{equation}
\widehat{V}=-\boldsymbol{H\cdot M}=-\frac{e}{2mc}\boldsymbol{H\cdot }\left[
\boldsymbol{r}\mathbf{\times }\boldsymbol{p}\right] ,
\end{equation}%
where $m$ is the bare electron mass that accounts for the
translational-invariant
symmetry breakdown due to the discrete crystal background \cite%
{Blount1962,Landau9}. (Analogous terms for the
time-symmetry-breaking field have been proposed in \cite{Manes} for
the orbital part of intrinsic spin-orbit coupling in graphene, which
is minute.)

The numerical parameter $\gamma $ of the matrix elements (\ref{gmatr}) is
estimated in the tight-binding nearest neighbor approximation (between sites
A and B) as:
\begin{equation}
\gamma \approx \frac{t/2}{{\hbar ^{2}}/{ma^{2}}}=0.4,
\end{equation}%
where $t=3.033$~eV is the $\pi $-$\pi $ hopping integral, and $a=1.42{%
\mathring{A}}$ is the hexagon side (C-C interatomic distance). It is more
difficult to estimate the next nearest neighbor parameter $\lambda $
(between A and A$^{\prime }$) in (\ref{lmatr}), since the hopping integral
falls off fairly slowly [$t_{\pi \pi }(d)\propto 1/d^{2}$] but, clearly, $%
\lambda <\gamma $.

Diagonalization of the Hamiltonian (\ref{HB}) can be easily done in terms of
LL wave functions $f_{n}(r)$, presenting the resulting 4-component
eigenfunctions $\tilde{F}^{n}(r)$ in the form (\ref{Flevels}) with slightly
different coefficients. This again gives the set of discrete LLs with $%
n=0,\pm 1,\pm 2,\dots $, having the energies:
\begin{equation}
E_{n}=s_{eh}\,\sqrt{2v^{2}|e|\hbar H_{z}|n|/c+\left( \gamma \pm \lambda
\right) ^{2}\mu _{B}^{2}H_{z}^{2}}.  \label{newE}
\end{equation}%
Special attention should be paid to zero LL $n=0$ with%
\begin{equation}
E_{0}=\left( -\lambda \pm \gamma \right) \mu _{B}H_{z},  \label{newE0}
\end{equation}%
and
\begin{equation}
\tilde{F}_{\pm }^{n}(r)=\{f_{0}(r),\;0,\;\pm f_{0}(r),\;0\}.
\end{equation}%
The new effect here, illustrated by Fig.~\ref{FigLevels}c, is the
valley-splitting of each LL, marked by the $\pm $ sign and estimated as:
\begin{gather}
\Delta E_{n}\simeq \gamma \lambda \left( {\frac{\mu _{B}^{3}H_{z}^{3}}{%
|n|mv^{2}}}\right) ^{\frac{1}{2}}\simeq 2\cdot 10^{-3}{\frac{\gamma \lambda
}{|n|^{1/2}}}H_{z}[\mathrm{T}]^{3/2}\,\mathrm{K,}  \label{Den} \\
\Delta E_{0}\simeq 2\gamma \mu _{B}H_{z}\simeq 1.3\gamma H_{z}[\mathrm{T}]\,%
\mathrm{K.}  \label{De0}
\end{gather}%
Being very small for non-zero LLs, $n\neq 0,$ this splitting should be
observable for zero LL in high fields. Note that this effect has purely
orbital origin and is completely decoupled from the additional Zeeman
spin-splitting shown in Fig.~\ref{FigLevels}d
\begin{equation}
\Delta E_{s}\simeq g\mu _{B}H\simeq 1.3{\frac{g}{2}}H[\mathrm{T}]\,\mathrm{K}%
,\quad g\approx 2,  \label{Des}
\end{equation}%
because of very weak spin-orbital coupling. Unlike orbital-splitting, the
spin-splitting is a function of the absolute value of $\boldsymbol{H}$
(rather than of $H_{z}$), that permits separating the two contributions $%
\Delta E_{s}$ and $\Delta E_{n}$ by their angular field dependence.

\begin{figure}[t]
\centering
\includegraphics [width=4cm] {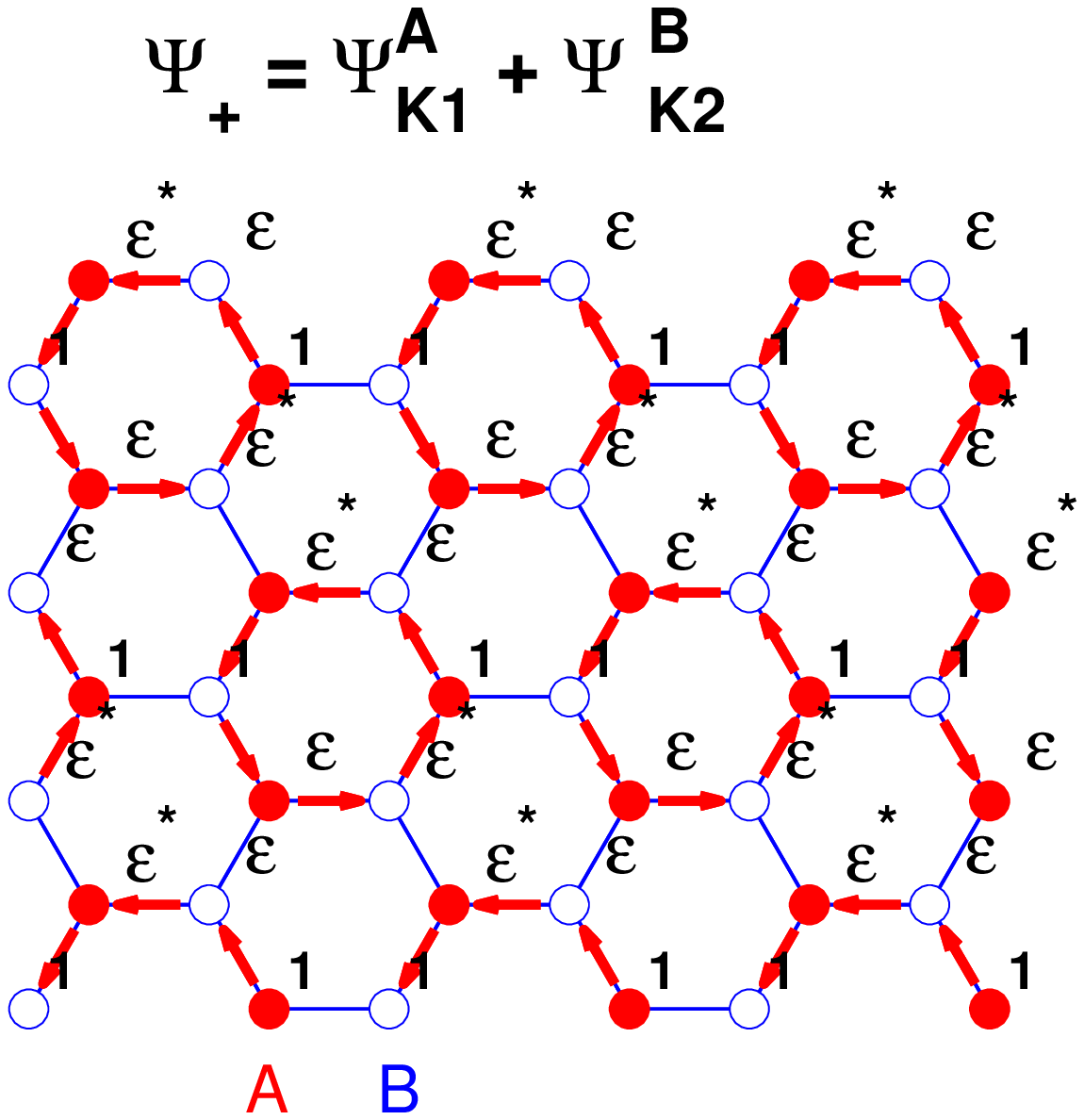} \ \includegraphics [width=4cm]
{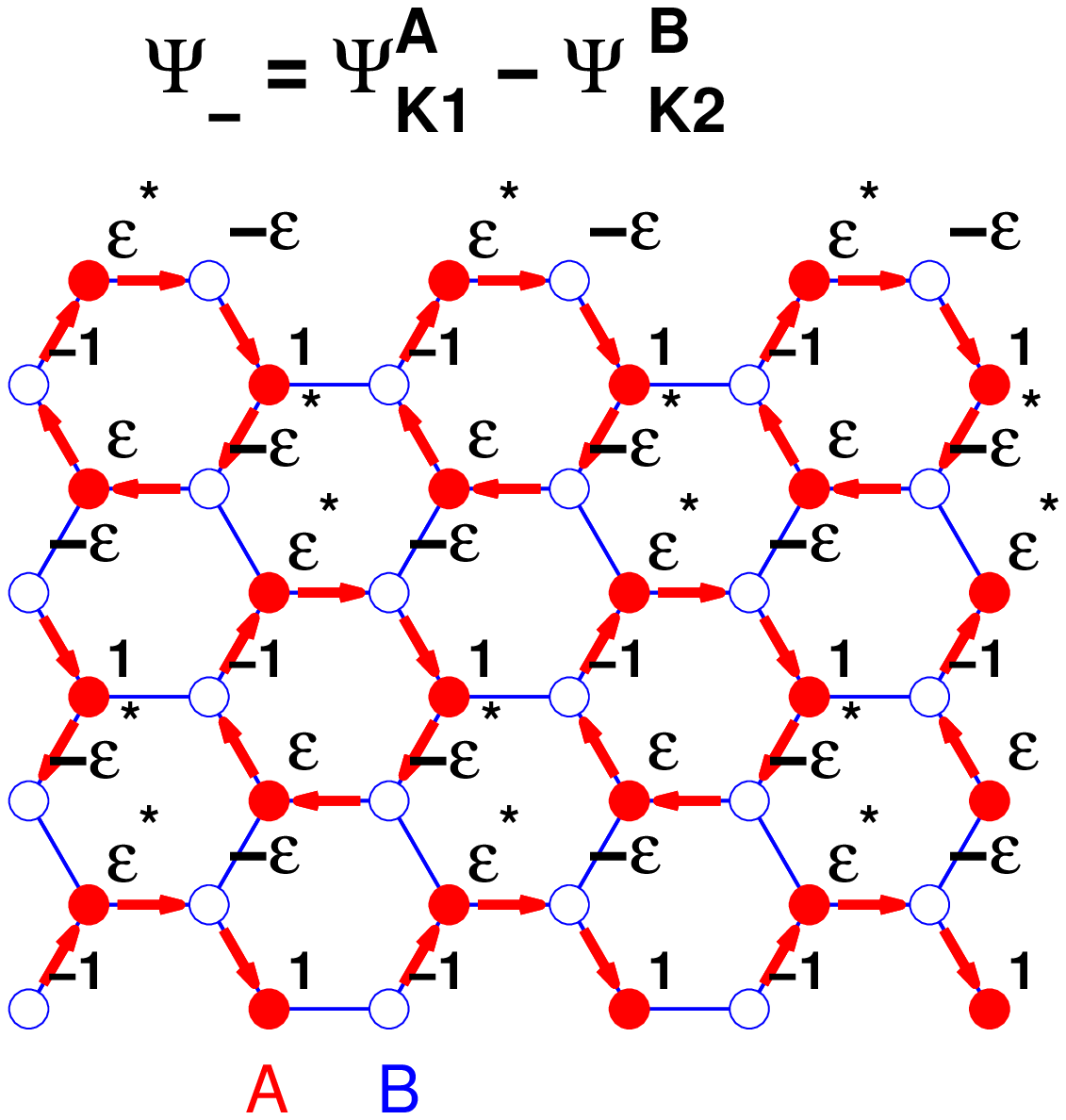}\newline
\caption{Schematic of circular currents corresponding to Bloch functions $%
\Psi _{\pm }=\Psi _{K_{1}}^{A}\pm \Psi _{K_{2}}^{B}$ of the split zero
Landau level.}
\label{FigCurrents}
\end{figure}

To clarify the physical origin of the orbital splitting, consider the
explicit form of the wave functions for carriers located on the up/down
shifted zero LL:
\begin{equation}
\Phi _{\pm }^{0}(r)=\sum_{i=1}^{4}F_{i\pm }^{0}(r)\Psi _{i}(r)=\Psi _{\pm
}(r)f_{0}(r),
\end{equation}%
where the Bloch parts
\begin{equation}
\Psi _{\pm }(r)=\ \Psi _{K_{1}}^{A}(r)\pm \Psi _{K_{2}}^{B}(r),
\end{equation}%
are presented in Fig.~\ref{FigCurrents} and can be interpreted as a set of
clockwise and counterclock-wise current loops circulating around every third
hexagon. Therefore, it is the orbital paramagnetic interaction of these
\emph{inter}-valley circular currents with $H_{z}$ that causes the splitting
$\Delta E_{0}$.

The current distribution for LLs with $n\neq 0$ is more complicated since
both clockwise and counterclock-wise current loops with different envelope
LL functions [$f_{n}(r)$ and $f_{n-1}(r)]$ contribute to the wave functions
of each split LL. The compensation of orbital momenta is almost complete,
which explains the negligibly small splitting (\ref{Den}) of higher LLs.
Note also that the additional contribution can be caused by the  \emph{intra}%
-valley circular currents circulating around next-nearest-neighbor
plaquettes proposed in \cite{Alicea2006}. Governed by the next-nearest
neighbor parameter $\lambda $, these currents do not contribute to $n=0$ LL
splitting and contribute only very weakly to the splitting of other LLs.
Another consequence of the orbital LL splitting is the lattice period
tripling produced by the network of circular current shown in Fig.~\ref%
{FigCurrents}. This field-induced breaking of graphene spatial symmetry can
be observed for non-integer filling of zero LL, when clock- and
counterclock-wise currents do not compensate each other.

To conclude, we have proved that the orbital mechanism is sufficient to
explain the zero LL splitting in graphene. The effect occurs in a
perturbative non-critical manner and is an intrinsic property of
noninteracting fermions on a hexagonal lattice. As a consequence (observable
optically), the orbital splitting should not depend on the LL filling
factor, unlike the result from other models. At the same time, the many-body
and/or disorder effects can amplify the orbital splitting (even for $n\neq 0$%
), induce an additional symmetry breaking, and bring about a nontrivial
field and filling factor dependence of the gap observed experimentally \cite%
{Zhang2006,Abanin2007}.

We grateful to Y. Kopelevich and M. Dyakonov for valuable discussions. IL
thanks the support of the ANR agency (project LoMaCoQuP).

\end{document}